\documentclass[referee]{aa}  
\usepackage[varg]{txfonts}
\usepackage{wasysym}
\usepackage{graphicx}
\usepackage{gensymb}
\usepackage{subcaption}
\usepackage{arydshln}
\usepackage{rotating}

\usepackage{txfonts}
\begin{document}

   \title{Validation scheme for solar coronal models - constraints from multi-perspective observations in EUV and white-light}


   \author{A. Wagner 
          \inst{1},
          E. Asvestari\inst{2}\fnmsep
          \and
          M. Temmer
          \inst{1}
          \and
          S.G. Heinemann
          \inst{1, 3}
          \and
          J. Pomoell
          \inst{2} }

   \institute{Institute of Physics, University of Graz,
              Universit\"atsplatz 5, A-8010 Graz\\
              \email{manuela.temmer@uni-graz.at}
         \and
             University of Helsinki\\
         \and
             Max-Planck Institut für Sonnensystemforschung, Justus-von-Liebig Weg 3, 37077 Göttingen, Germany\\
             }

   \date{Received June xx, 2021; accepted xx}

 
  \abstract
   {In this paper we present a validation scheme to investigate the quality of coronal magnetic field models, which is based upon comparisons with  observational data from multiple sources.}
   {Many of these coronal models may use a range of initial parameters that produce a large number of physically reasonable field configurations. However, that does not mean that these results are reliable and comply with the observations. With an appropriate validation scheme the quality of a coronal model can be assessed. }
   {The validation scheme is developed on the example of the EUropean Heliospheric FORecasting Information Asset (EUHFORIA) coronal model. For observational comparison we use extreme ultraviolet and white-light data to detect coronal features on the surface (open magnetic field areas) and off-limb (streamer and loop) structures from multiple perspectives (Earth view and the Solar Terrestrial Relations Observatory - STEREO). The validation scheme can be applied to any coronal model that produces magnetic field line topology.}
   {We show its applicability by using that validation scheme on a large set of model configurations, which can be efficiently reduced to an ideal set of parameters that matches best with observational data.}
   {We conclude that by using a combined empirical visual classification with a mathematical scheme of topology metrics a very efficient and rather objective quality assessment for coronal models can be performed.}


   \maketitle
%

%

\section{Introduction}

The solar wind and embedded structures, such as coronal mass ejections (CMEs) and high speed streams (HSS) are key components of Space Weather, and thus of great interest to the space weather forecasting community. Simulations generating accurate reconstructions of the solar wind structure are necessary for studying the propagation behavior of CMEs and their interaction processes with the ambient solar wind \citep[see][]{schmidt01,case08,temmer11,sachdeva15}. Moreover, such reconstructions provide the basis for reliable Space Weather alert systems, in terms of forecasting CME arrival time and speed as well as high speed solar wind streams. Currently, for simulating the inner heliosphere, a plethora of heliospheric propagation models are available, both empirical and full magnetohydrodynamic (MHD) ones \citep[see e.g.,][and references therein]{riley18, temmer21,vrsnak21}. The majority of such models require accurate boundary conditions, namely coronal magnetic field structure and plasma properties, provided at a few solar radii away from the Sun. These boundary conditions are produced by coronal models, and thus, the efficiency of the latter strongly influence the quality of the heliospheric model results. Subsequently, the assessment of the quality of coronal models is necessary for interpreting simulation results of the inner heliospheric solar wind structure and propagating transients. For model development as well as further improvement of up-to-date Space Weather tools, we need rigorous evaluation of basic coronal model performances close to the Sun, in addition to planetary targets \citep[see e.g.,][]{Hinterreiter19, Sasso19}. Currently no systematic validation procedures for coronal models are available, apart from individual studies \citep[e.g.,][]{Cohen07, Jian16, Yeates18, Meyer20}, some of which often focus on a single coronal model
and only one or two input parameters \citep[see for example][]{Asvestari19}. 


As for any model, the choice of parameter settings can be plentiful, which leads to a large variety of physically meaningful solutions. Moreover for the coronal model, the only observational input, the magnetogram, appears to have significant effects on the model results too \citep[e.g.,][]{Riley2014, linker21}. Therefore, the quality of each solution needs to be validated and quantified in its reliability in order to derive an optimum set of model parameters. For restricting and better understanding the choice of input parameter values, we present an objective validation scheme, that can be used for any coronal model which provides results for the magnetic field line topology. The validation scheme is tested on the up-to-date numerical coronal model part of EUropean Heliospheric FORecasting Information Asset (EUHFORIA) \citep{PomoellPoedts18} covering distances from close to the Sun up to 21.5~R$_{\odot}$ (0.1~AU). 

The methodology we developed is based on matching simulations with observations for off-limb features over various distances observed in white-light, and on-disk open and closed magnetic field areas observed in extreme ultraviolet (EUV) frequencies. Plumes, fans, (helmet) streamers, and large-scale loops are tracers of open and closed magnetic field structures, making them ideal for coronal model evaluation. Moreover, coronal streamers are assumed to be one of the slow solar wind sources \citep{sheeley97,cranmer17}, and therefore are of special interest when comparing with coronal model results. Using remote sensing image data, the plane of sky projected signatures of those features appear differently when viewed from multiple viewpoints. Therefore, to obtain a more clear picture of the three dimensional features it is important to use white-light observations from different vantage points. For that, we employed observational data that include images both from \emph{Solar and Heliospheric Observatory}/Large Angle and Spectrometric COronagraph \citep[SOHO/LASCO,][]{Brueckner1995} as well as enhanced solar eclipse photographs produced by Druckmüller and \emph{Solar TErrestrial RElations Observatory}/Sun Earth Connection Coronal and Heliospheric Investigation \citep[STEREO/SECCHI,][]{Howard2002, Kaiser2005}, providing us with the advantage of investigating the effects of projection. Using these we identify high-quality model results which are matching simultaneously observations from various viewpoints. Applying the methodology on two benchmark dates (01-Aug-2008 and 11-Jul-2010, both dates of a total solar eclipse), we assess and quantify the model quality for each parameter set.

In Section \ref{sec2} we first describe the model specifics of EUHFORIA's coronal model as well as the observational data that was used in this exemplary study. We then present the methodology of the validation scheme in Section \ref{sec3}. The results of the analysis itself are shown in Section \ref{sec4}, followed by a discussion and conclusion of the outcomes in Section \ref{sec5}.


\section{Coronal model and observational data}
\label{sec2}
\subsection{Coronal Model description}

EUHFORIA is divided into two modeling domains, the 'coronal domain' and the 'heliospheric domain'. The coronal domain consists of a PFSS \citep[Potential Field Source Surface,][]{Arge2000} model, coupled with an SCS \citep[Schatten Current Sheet,][]{Schatten1971} model \citep{PomoellPoedts18}. The PFSS model computes the magnetic field configuration up to the source surface height $R_{\rm ss}$ from a scalar potential, thus, assuming the domain of calculation to be current-free. All modeled field lines that are anchored at both ends in the photosphere are designated as closed. However, those field lines that extend above it are considered to be open field lines, and thus are the ones that contribute to the interplanetary magnetic field (IMF). In terms of the modeling domain above this $R_{\rm ss}$ height the majority of magnetic field lines are extended radially up to the domain boundary at 0.1~AU, while in addition some field lines bend from higher to equatorial (low) latitudes. The current-free assumption considered in the lower corona for the PFSS model is a rather inaccurate assumption for the upper corona, as expanding the field lines purely radially would create a rather broad heliospheric current sheet. 
Thus the SCS model is coupled with the PFSS to model the magnetic field topology beyond $R_{\rm scs}$, which then also incorporates $B_\theta$ and $B_\phi$ components to reproduce the observed thin current sheet. To avoid discontinuities between the models at that boundary, the so-called Schatten Current Sheet height $R_{\rm scs}$, which is the inner boundary of the SCS model, is placed below the source surface height $R_{\rm ss}$ \citep[see][]{mcgregor08,Asvestari19}. The EUHFORIA coronal modeling domain was calculated with a mesh grid having a resolution of 0.5 degrees per pixel, while for solving the Laplacian equations for the PFSS and SCS calculation, solid harmonics up to the order of 140 were used.

Considering that the only requirement is that $R_{\rm scs}<R_{\rm ss}$, a variety of possible height values and their combinations exist, usually covering distances of about 1.2--3.25~R$_{\odot}$. For our purpose the EUHFORIA coronal model is initiated with 67 different parameter sets covering the boundary heights $R_{\rm ss}$ within 1.3--2.8~R$_{\odot}$ and $R_{\rm scs}$ within 1.4--3.2~R$_{\odot}$ \citep{Asvestari19}. We produce results for the full 3D configuration of the magnetic field for all 67 parameter sets that are visualized as field lines applying the visualization software VisIt \citep{VisIt}. The PFSS solution is plotted in a 3D sphere, where the field lines are traced outwards with their starting points being distributed on a uniform grid in longitude and latitude on the solar surface. The SCS solution is shown as a 2D slice of field lines uniformly distributed in latitude in the plane of sky. For comparison and validation with off-limb features, we compute the open and closed magnetic field areas on the solar surface and overplot the simulated field lines from the corresponding viewing angles onto white-light images.

\subsection{Observational Data description}

The dates selected for this study are 01-Aug-2008 and 11-Jul-2010, respectively, as these are both eclipse dates, and thus additional ground-based imagery data of the fine structures of the solar corona are available. Observational input for modeling the solar corona traditionally comes from magnetograms, measuring the magnetic field configuration in the photosphere. For 01-Aug-2008 we use the synoptic magnetic field map from GONG \citep[Global Oscillation Network Group;][]{Harvey1996} and for 11-Jul-2010 the synoptic map produced by 720s-HMI \citep[Helioseismic and Magnetic Imager][]{Schou2012, Couvidat2016} aboard SDO \citep[Solar Dynamics Observatory;][]{Pesnell12}. For the comparison of the model results with observations, we use for the off-limb structures white-light data from SoHO  \citep[Solar and Heliospheric Observatory;][]{Domingo1995} and both STEREO-A/B \citep[Solar Terrestrial Relations Observatory;][]{Kaiser08} satellites. The multiple spacecraft data increase the statistical samples for comparison and moreover, enable us to compare the model results with simultaneous observations from three different viewing angles. Furthermore, we make use of the availability of high-resolution solar eclipse images by Druckmüller\footnote{\url{http://www.zam.fme.vutbr.cz/~druck/eclipse/}} using sophisticated image processing techniques \citep{Druckmueller06,druckmueller09}. A clear advantage of those eclipse images over other image data is that even rather faint coronal structures are unveiled starting from the solar limb up to several solar radii. While the STEREO SECCHI/COR1 \citep{Howard08} instruments with a field of view (FoV) from 1.5 to 4~R$_{\odot}$ serve the purpose of comparing structures close to the limb, LASCO-C2 \citep[FoV: 2.2 to 8
~R$_{\odot}$;][]{Brueckner1995} and COR2 (FoV: 2.5 up to 15~R$_{\odot}$) are used for the comparison of the outer parts of the modeling domain. The eclipse data were used for both purposes. 

For STEREO COR1 imagery we apply a normalizing radial graded filter (NRGF) processing technique \citep[see][available under IDL SolarSoftWare]{Morgan06} and additional contrast enhancement to improve the visibility of streamers further away from the Sun. In addition, images within a 20 minute window were stacked. No such procedures were applied for the LASCO-C2 data as visibility of features and general contrast were sufficient for the analysis.

For the comparison of model results with observed open magnetic field regions on the solar disk (i.e., coronal holes), we use synoptic image data from SOHO/EIT 195\AA~and SDO/AIA 193\AA~by \cite{HessWebber14} and \cite{Karna14}, respectively. The resolution is set to 0.5 degrees per pixel.

\section{Validation methods}
\label{sec3}
In the following we present validation algorithms that are employed to quantify the quality of model results and parameter sets that were used. Each method can be used on its own, but most efficiently they are used in combination with a specific workflow. The methods cover a very basic visual inspection (Section~\ref{sub:visual}) as well as sophisticated metrics that quantify the matching of the morphology of off-limb features (Section~\ref{sub:feature}) and open/closed magnetic field on the Sun (Section~\ref{sub:topology}). We first describe the stand-alone methods and then present, based on the two selected dates, the developed workflow and results (Section~\ref{sec4}).

\subsection{Method I: Visual Classification}\label{sub:visual}
\begin{figure} 
    \centering
    \includegraphics[width=0.7\linewidth]{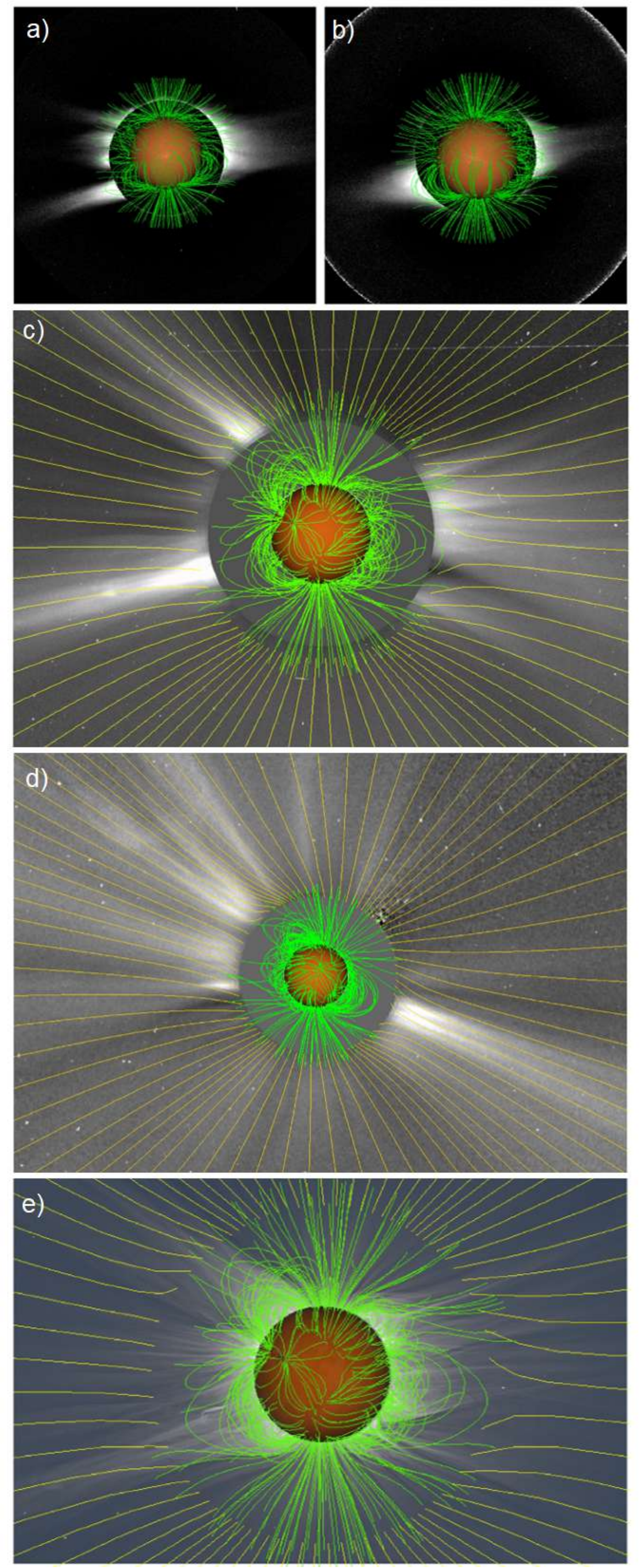}
    \caption{Showcase of visualization of field lines (uniformly sampled) from the model, overplotted to observational white-light data from: a) STEREO B COR1, b) STEREO A COR1, c) SOHO LASCO C2, d) STEREO A COR2 and e) eclipse picture. The PFSS model is plotted as 3D field line configuration in green, while the SCS solution is plotted in a 2D plane-of-sky slice in yellow. In a) and b) $R_{\rm ss} = 2.4 R_\odot$ for 01-Aug-2008, while in c), d) and e) $R_{\rm ss} = 2.9 R_\odot$ and $R_{\rm scs} = 2.5 R_\odot$ for 11-Jul-2010.}
    \label{Fig Visual Showcase}
\end{figure}

\begin{figure} 
    \centering
    \includegraphics[width=0.85\linewidth]{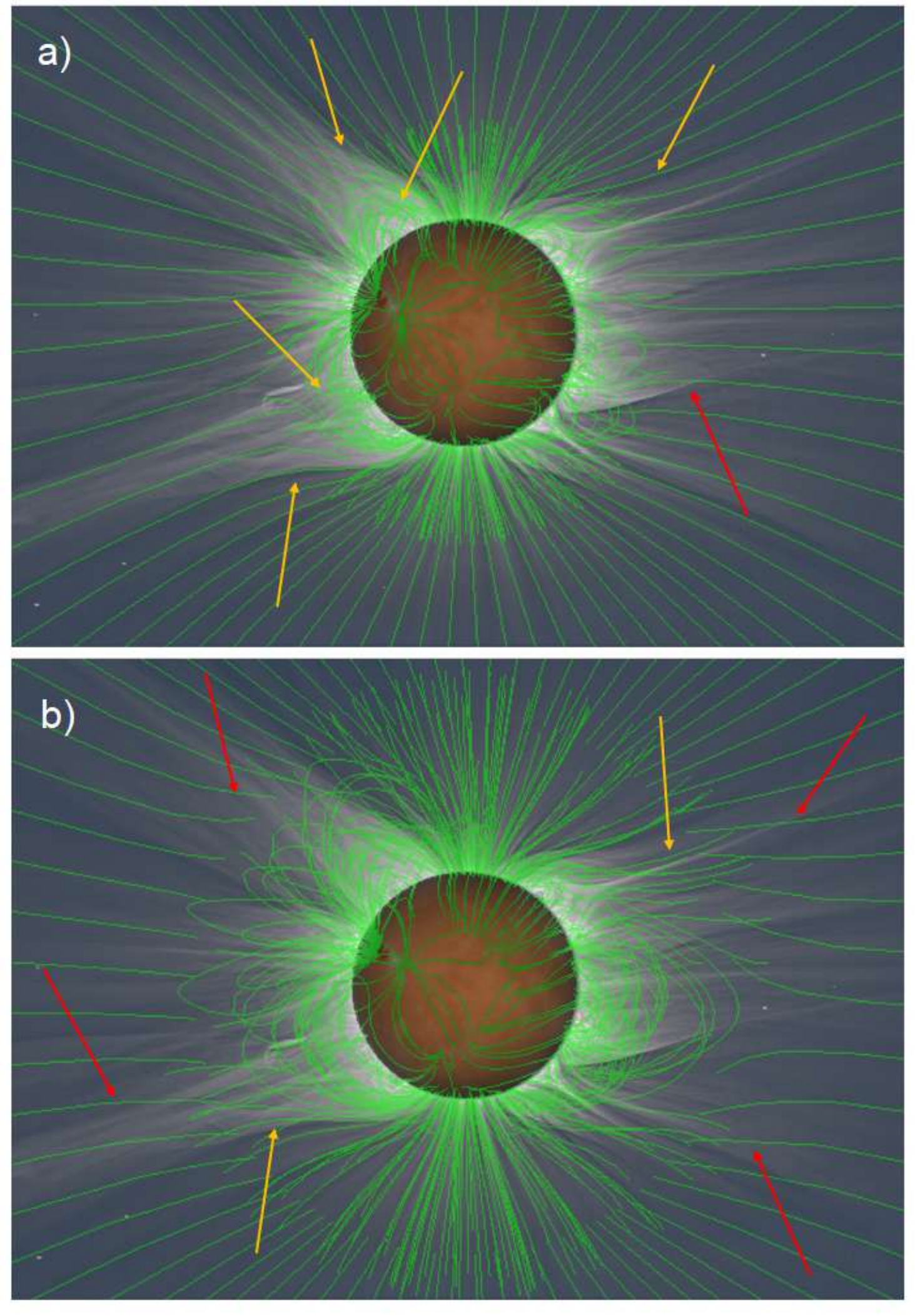}
    \caption{Illustration of the visual classification process for two configurations on 11-Jul-2010 overlaid on eclipse data. Yellow arrows mark good matches between observation and field line simulation, while red arrows mark mismatches. Panel a) thus shows a well matching configuration ($R_{\rm ss} = 1.9 R_\odot$ and $R_{\rm scs} = 1.5 R_\odot$), while the field line solution in b) is rejected by our criteria ($R_{\rm ss} = 2.8 R_\odot$ and $R_{\rm scs} = 2.4 R_\odot$).}
    \label{Fig Visual2}
\end{figure}

Though less objective, the visual classification is an efficient method to quickly assess the quality of modeling results. A simple overplot of model results on white-light coronagraph data is used to roughly distinguish between high and low quality results by inspecting the agreement between observed white-light features and modeled field lines (see Figure \ref{Fig Visual Showcase}). Since we use the coupled PFSS+SCS model, we use the visual classification especially for an empirical assessment of the field line behavior at the boundary between the two model domains, namely, the field line bending of the SCS model at lower heights. Stronger constraints using this simple method can be given by adding data from multiple viewpoints as provided by SoHO-LASCO/C2 and STEREO-SECCHI/COR2. Fine structures, showing in more detail, for example, the bending of field lines, are obtained by using eclipse image data. In Figure \ref{Fig Visual Showcase} the configurations in panels a) and b) show a matching of the loop structures with the overlying bright features in the coronagraph COR1 of STEREO B and A, respectively. On the other hand, in panels c), d) and e) we lay our focus on the field line bending of the SCS close to the source surface, where we can see mismatches across all panels between the edges of the bright structures in white-light with the field line trajectories close to them. Though the visual classification method is rather subjective, for most configurations a clear distinction between match and mismatch can be derived as exemplary given in Figure \ref{Fig Visual2}.

\subsection{Method II: Feature Matching}\label{sub:feature}
\begin{figure*}
    \centering
    \includegraphics[width=0.85\linewidth]{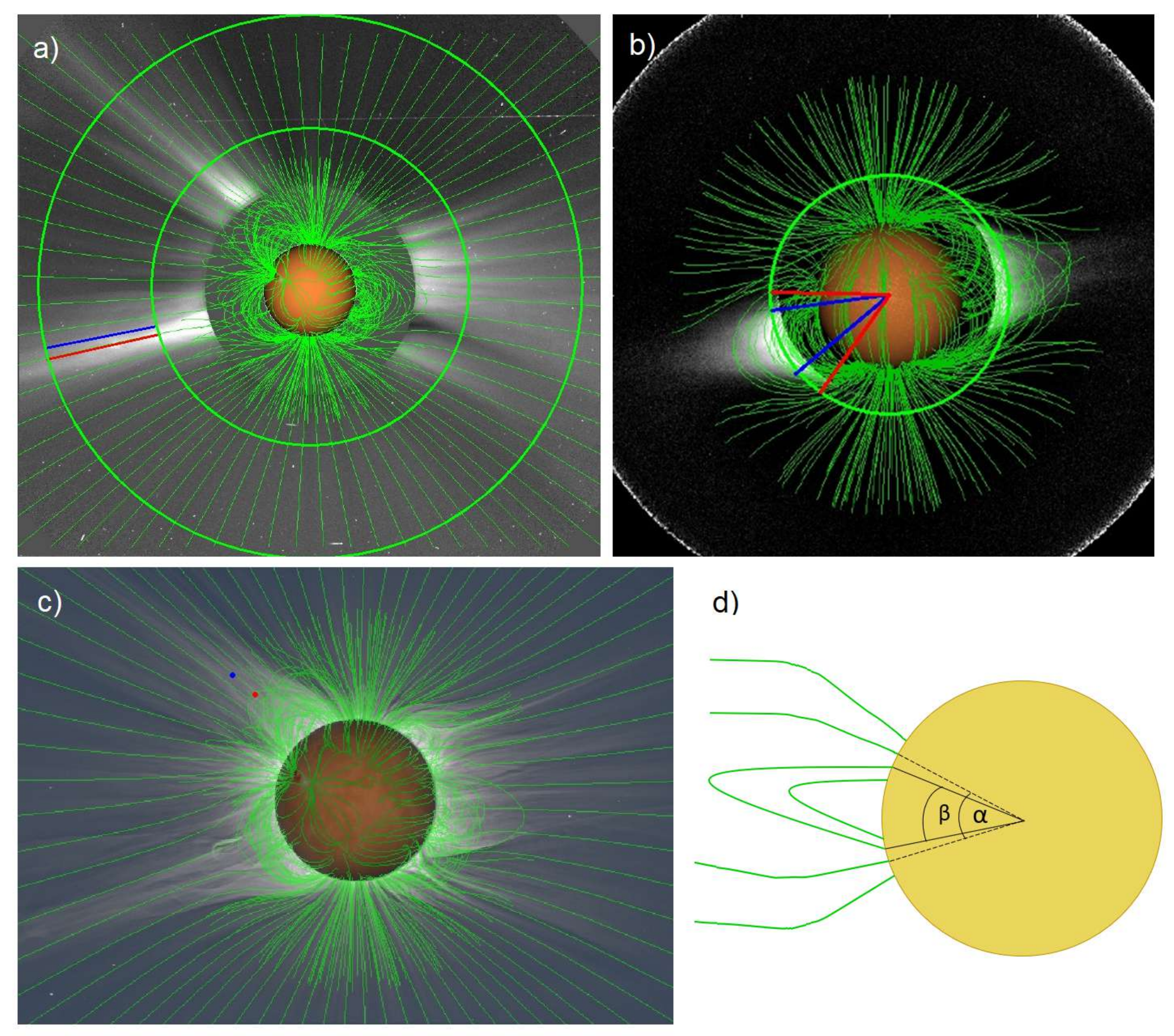}
    \caption{Chosen features and visualization of feature matching methods with a) streamer direction method from II.a (using LASCO C2 for 11-Jul-2010, $R_{\rm ss} = 2.9 R_\odot$ and $R_{\rm scs} = 1.9 R_\odot$), b) streamer width method from II.a (STEREO A COR1 for 01-Aug-2008, $R_{\rm ss} = 3.0 R_\odot$), c) Brute Force matching method II.b (using an high resolution eclipse image by Druckmüller for 11-Jul-2010, $R_{\rm ss} = 2.4 R_\odot$ and $R_{\rm scs} = 1.4 R_\odot$). In  a), b) and c), blue colored marking result from the observation, while red marking result from the modelled field lines. d) gives an illustration of the two possible definitions of the streamer width by the underlying closed topology of the model (closed field - angle marked by solid lines; first open field - angle marked by dashed lines). }
    \label{Fig: FM}
\end{figure*}

In comparison to the visual inspection described in Method I, a semi-automatized identification of matching white-light features to modelled ones provides a more objective, but more time-consuming (in terms of human intervention), method that results in a quantitative assessment of the quality of model results. For comparing different features between model and observational data in an efficient way, we use simple point-and-click algorithms. As there are many possibilities, here we opted to compare in Section \ref{ssec3_2_1} streamer orientation angles with the SCS field line directions as well as opening angles (width) of streamers with the boundary of closed to open topology in the PFSS, and in Section \ref{ssec3_2_2} the Brute Force feature matching to identify differences in the location of certain structures (see Figure~\ref{Fig: FM}).

\subsubsection{Method II.a: Streamer direction and width}
\label{ssec3_2_1}
Coronal streamers are quasi-static features that are shaped by the global magnetic field structure of the Sun and appear as bright structures in coronagraphs, and thus are well-observed without intensive processing of image data. Hence, they are well suited for a comparison with models showing the global coronal magnetic field. Helmet streamers are located above regions of closed field lines, such as active regions or filament channels, having a certain extension (width) and thinning out into a ray-like structure with a rather radial orientation enveloping a current sheet \citep[see for example the review by][]{koutchmy92}. Similar in appearance but without a current sheet are unipolar pseudo-streamers, connecting two coronal holes of same polarity \citep[see][]{wang07}.

For the streamer direction we use the SCS model results which are visualized in 2D slices from which we derive the orientation of field lines in an image plane. We assume that the brightest streamer is lying closer to the image plane, hence, it is the most ideal to use for comparing the modelled field in the outer corona to coronagraph images from SOHO/LASCO-C2 and COR2 aboard STEREO-A and -B. We measure the angular difference between the modelled field line and observed streamer orientation over the heights $H_1 = 3.5 R_{\odot}$ to $H_2 = 6.0 R_{\odot}$. The values were chosen such that the model field lines as well as the streamer structure are approximately radial but still well visible in the coronagraph FoV. In the showcase in panel a) of Figure \ref{Fig: FM} we see that the directions of both the tracked streamer edge (blue), as well as the marked field line (red) match quite nicely for this configuration with the angular difference being only 0.5 degrees, which is well within the uncertainties of the method (cf. Figure \ref{FM1}).

The width of the streamer base-structure is observed in the low corona and can be detected from COR1 STEREO-A and -B white-light data over 1.5--4~$R_\odot$. That distance range can be applied for validating modelled coronal magnetic field structures in the PFSS domain of the model. At a fixed height above the photosphere, which was chosen to be $H = 1.75 R_{\odot}$, we measure the streamer width in the image data and the loop extensions in the model results. An example can be seen in Figure \ref{Fig: FM} in panel b). Here the extension of the modelled loop structures surpasses the width of the white-light feature at $H = 1.75 R_{\odot}$ substantially with 48.9 degrees for closed fields as boundary and 55.3 degrees for open fields as boundary compared to 34.8 degrees from the white-light image. It thus shows a poorly matching configuration for this sub-step. The height is chosen so that it is well above the occulting disk of the coronagraph, in order to avoid stray light effects, but also as low as possible in order to capture the loop extension for as many model configurations as possible, especially for those with a low source surface height.

\subsubsection{Method II.b: Brute Force Feature Matching}
\label{ssec3_2_2}
In comparison to coronagraph data, contrast enhanced solar eclipse images cover the coronal fine structures better due to the Moon being almost an ideal occulter. That enables to investigate features over large distance ranges with high accuracy. Subsequently, this can be used for a more detailed comparison using feature matching methods such as Brute Force, where the positions of features that are suspected to be the same are compared directly via point-and-click. In principle this provides a large variety of possibilities and is therefore a rather flexible approach of comparing with certain features of a model. Here, shown in panel c) of Figure~\ref{Fig: FM}, we compare with that method the apex location of loop systems, as it can be well identified from observational data and from model results. The method has no fixed heights and selects the most well observed features.

\subsection{Method III: Topology Classification}\label{sub:topology}
To quantify the quality of the model results in a fully automatized and objective way, we investigate the magnetic field topology. We assume that the majority of the open field emanates from coronal holes, which are usually observed in EUV as structures of reduced emission \citep{Cranmer2009}, due to reduced density and temperature in contrast to the surrounding quiet sun \citep[e.g.,][]{Heinemann2021}. As the coronal model covers the Sun over 360 degrees, we use for the extraction synoptic EUV maps. Checking the bimodal logarithmic intensity distribution and visually identified boundaries of coronal holes \citep[e.g.,][]{krista09,rotter12} we use as threshold $log(EIT_{\rm data})=2.95$ for 01-Aug-2008, and $log(AIA_{\rm data})=3.5$ for 11-Jul-2010. The extracted coronal hole areas are compared to the computed open field regions from the coronal model. The areas outside coronal holes are assumed to be closed field and are also compared with those from the model results. The modelled maps of magnetically open and closed regions are compared with the EUV maps, scaled to the size of each other, by applying three different metrics: 

1) The coverage parameter, $$P_{\rm cov} = \frac{N_{\rm o,b}}{N_{\rm o,EUV}}*100 \%,$$ where $N_{\rm o,b}$ is the number of pixels that are found to be open in both maps, EUV and model, and $N_{\rm o,EUV}$ is the number of open pixels in the EUV map,

2) the Jaccard metric for open fields, $$P_{\rm Jac} = \frac{N_{\rm o,b}}{N_{\rm o,all}}*100 \%,$$ with $N_{\rm o,all}$ representing all pixels that are open in either EUV or the model, and

3) the global matching parameter, $$P_{\rm glob} = \frac{N_{\rm match}}{N_{\rm tot}}*100 \%,$$ with $N_{\rm match}$ the number of pixels where the topology, either open or closed, matches in both maps and $N_{\rm tot}$ the total number of pixels. To avoid misinterpretation due to the large uncertainties coming from the polar regions, we cut the maps to heliographic latitudes in [$-$60, $+$60] degrees and count pixels only within that range. 
  
$P_{\rm cov}$ was already used to quantify EUHFORIA's accuracy to model coronal hole areas in \citet{Asvestari19}, and gives the fraction of overlap between modelled and observed open regions. $P_ {\rm Jac}$ expresses where model results produce open magnetic field topology which is not observed in EUV, and $P_{glob}$ defines the overall correctly modelled topology fraction.


\section{Results}
\label{sec4}
\begin{figure}
    \centering
    \includegraphics[width=\linewidth]{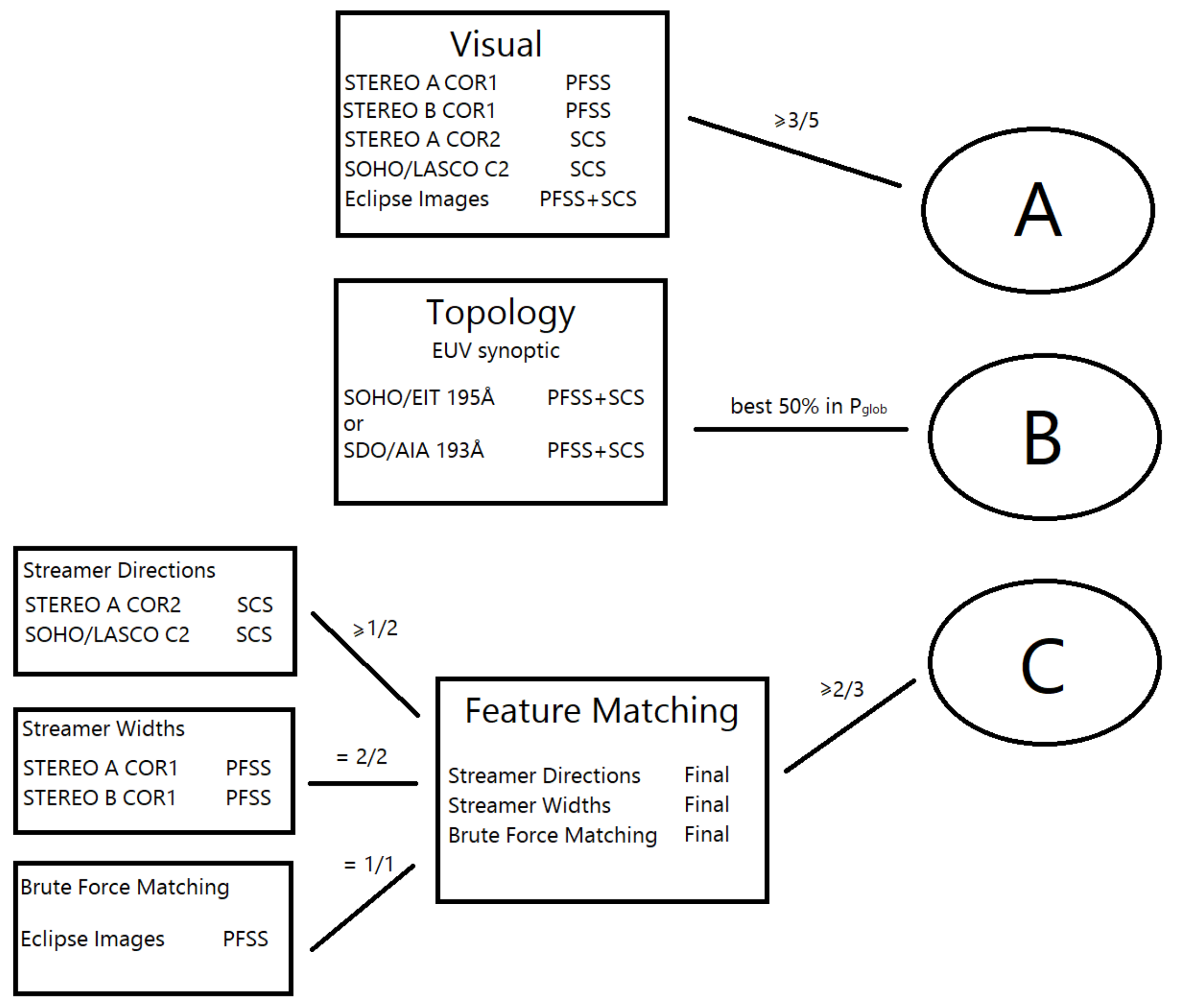}
    \caption{Workflow of our application of the benchmarking system on the EUHFORIA coronal model. In each box on the right there are the comparative images - except in the "Feature Matching" box, where there are the results of each sub-step on the right. Configurations that passed the analysis given in the boxes are sorted into the sets A, B and C.}
    \label{Fig: Workflow}
\end{figure}

The stand-alone methods as described in Section~\ref{sec3}, are most efficient when combined in the frame of a certain workflow. On the basis of EUHFORIA's coronal model for the two selected dates 01-Aug-2008 and 11-Jul-2010, we present, in the following, a developed sequence of empirical classification and physical/mathematical methods for quantifying and validating the coronal model results. That workflow is depicted in Figure \ref{Fig: Workflow} and we describe below the application from top to bottom. The full set of model parameters covering a total of 67 different configurations as well as the selected subsets (A, B, C) are given in the Appendix in Table \ref{conf_table}.

We first carried out the visual inspection of off-limb structures (see Section \ref{sub:visual}) starting from the full set of 67 model configurations. Observational data from five different sources - STEREO-A COR1, STEREO-B COR1, STEREO-A COR2, LASCO-C2, eclipse image are overlaid with the computed magnetic field topologies. We note that STEREO-B COR2 images show for both dates only low-intensity structures, hence, are not used for further analysis. We then visually inspect these images and check the general match with the observations for PFSS in the lower corona and the field line bending as derived from the SCS model. We obtain, on the one hand, that the larger values of SCS boundary height produce field line bendings at distances where streamers are observed to be already mostly radial. On the other hand, the low end of the $R_{\rm scs}$ parameter values spectrum shows a strictly radial behavior of field lines where bending can still be seen in observational data. Hence, we may restrict our parameter set so that heights in the lower mid of the parameter spectrum of the SCS model are preferred. For both dates, 01-Aug-2008 and 11-Jul-2010, the best visual match is found in the interval $R_{\rm scs} \in [1.5; 2.1] R_{\odot}$. If three out of five images show a good visual match with the model results from the different perspectives we keep that model configuration and form a selection set A (cf., Figure \ref{Fig: Workflow}) consisting of 30/32 parameter sets for 01-Aug-2008/11-Jul-2010.

\begin{figure}
     \centering
     \includegraphics[width=1\linewidth]{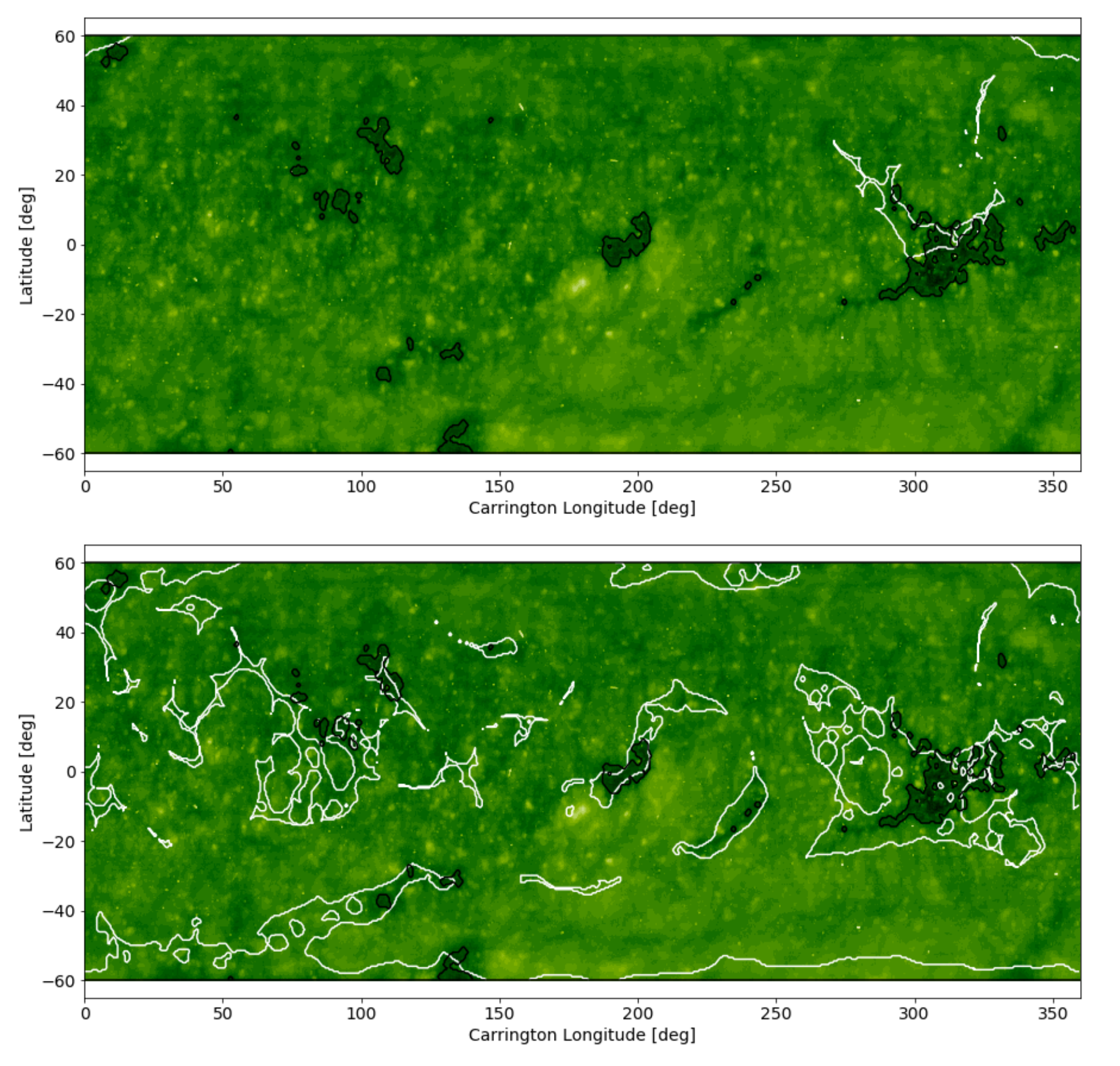}
     \caption{EIT EUV Carrington map for 01-Aug-2008 with the extracted open areas outlined in black. Open fields computed with EUHFORIA (white outlines) for the configuration of $R_{\rm ss} = 3.2 R_\odot$ and $R_{\rm scs} = 2.8 R_\odot$ (top) as well as the configuration of $R_{\rm ss} = 1.4 R_\odot$ and $R_{\rm scs} = 1.3 R_\odot$ (bottom).}
     \label{Fig Topology Maps 2008}
\end{figure}

\begin{figure}
     \centering
     \includegraphics[width=1\linewidth]{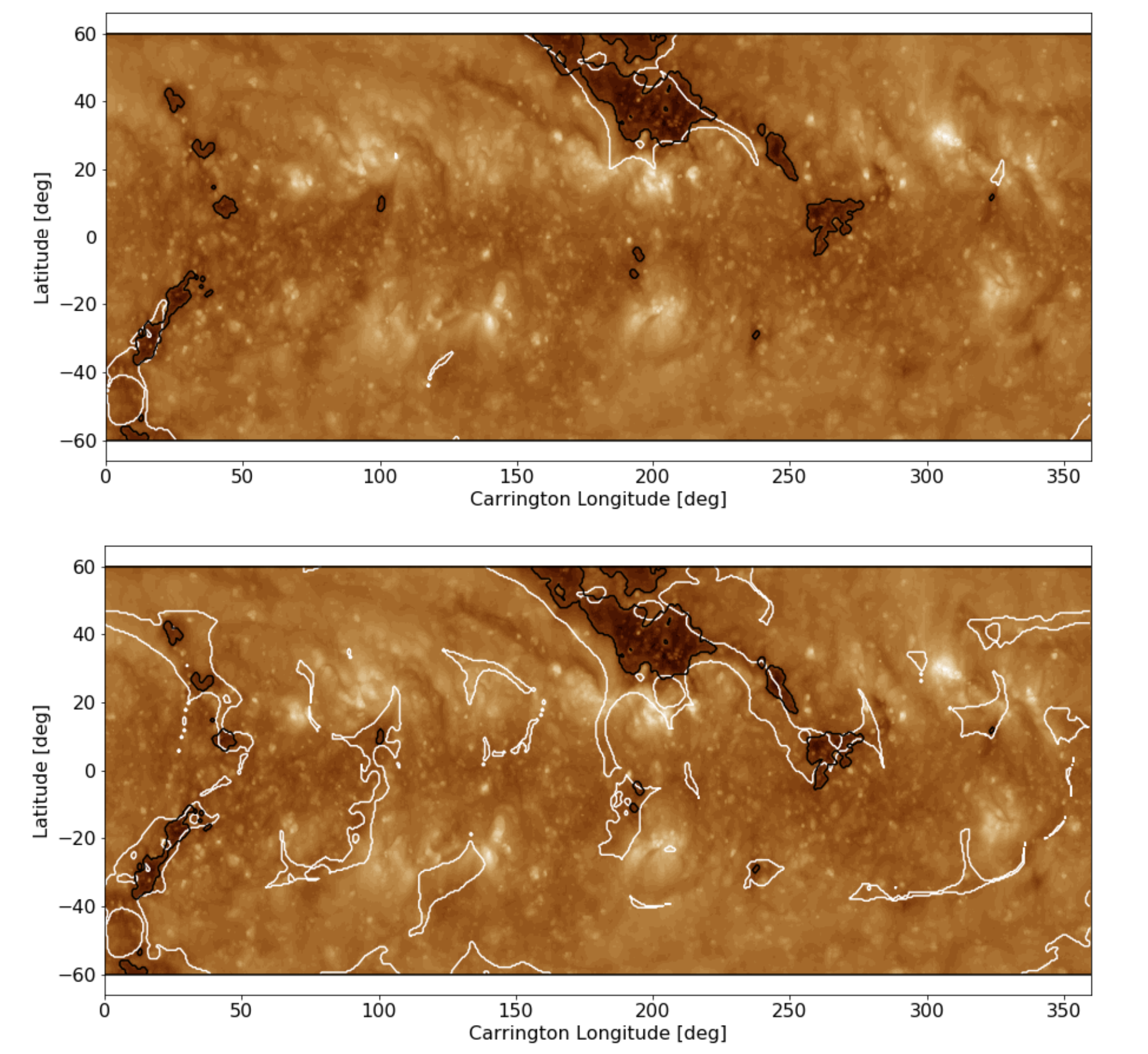}
     \caption{The same as Figure~\ref{Fig Topology Maps 2008} but for the AIA EUV Carrington map for 11-Jul-2010.}
     \label{Fig Topology Maps}
\end{figure}

\begin{figure*}
    \centering
    \includegraphics[scale = 0.25]{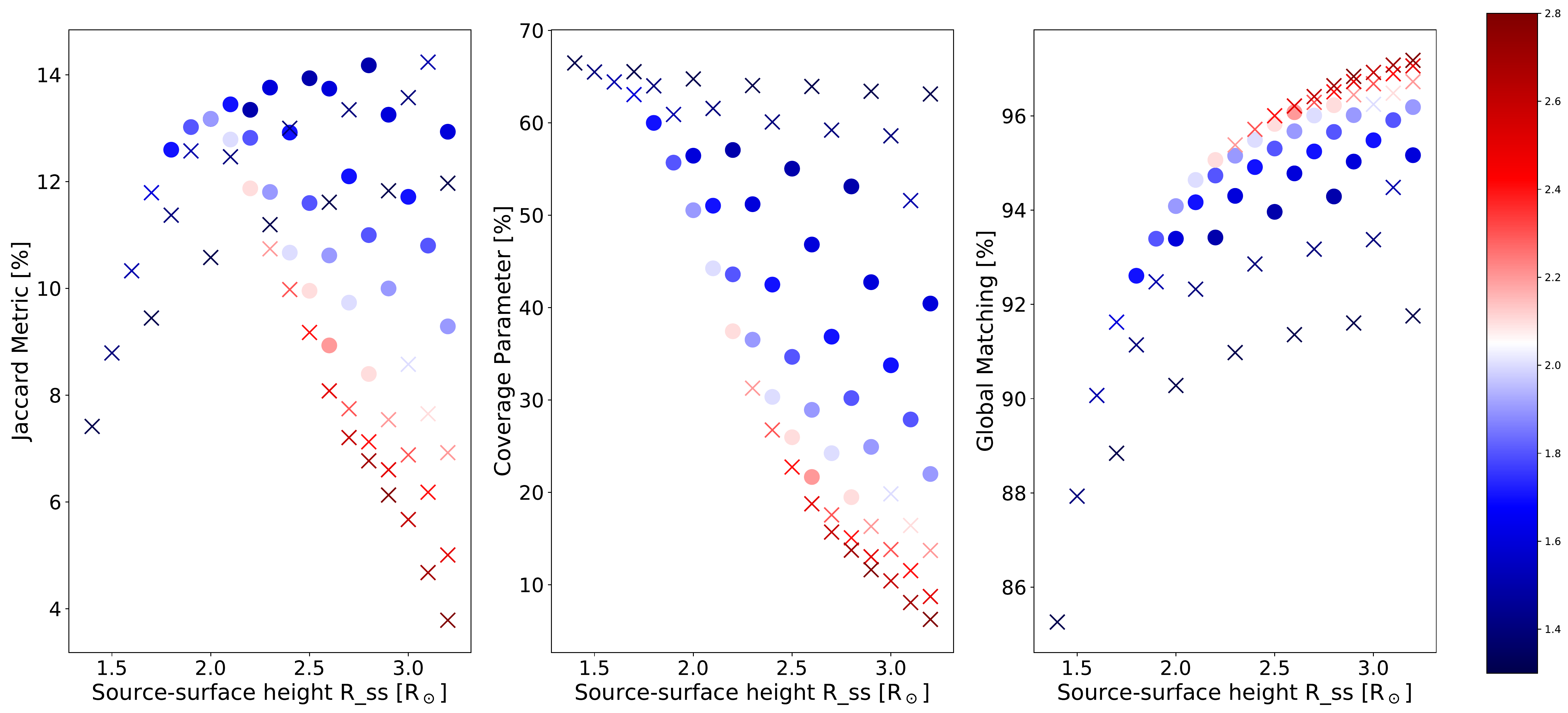}
    \caption{Behavior of $P_{\rm Jac}$, $P_{\rm cov}$ and $P_{\rm glob}$ with varying PFSS and SCS heights for 01-Aug-2008. The color bar indicates the SCS heights, crosses mark configurations that failed in the visual classification while dots mark configurations that passed it.}
    \label{Fig Topology 2008}
\end{figure*}

\begin{figure*}
    \centering
    \includegraphics[scale = 0.25]{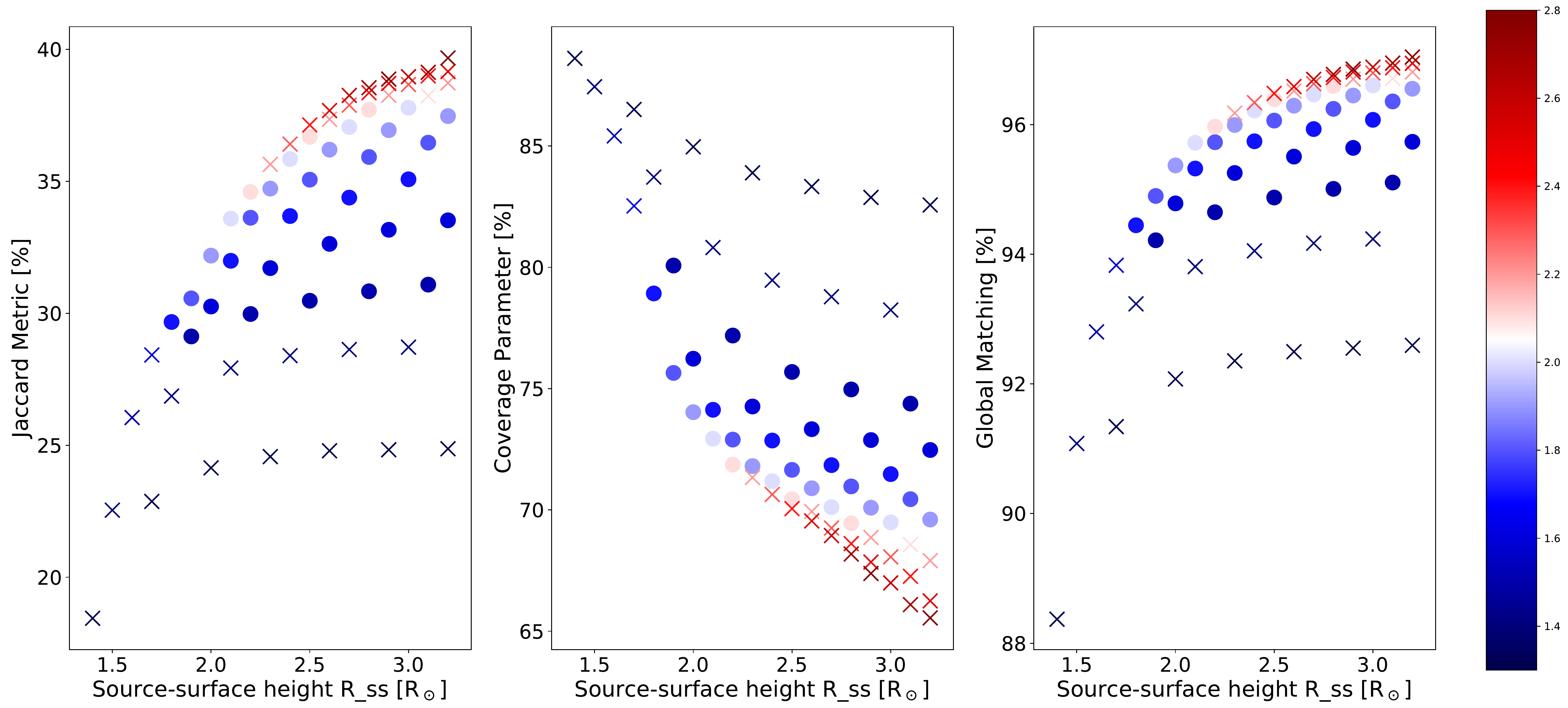}
    \caption{Behavior of $P_{\rm Jac}$, $P_{\rm cov}$ and $P_{\rm glob}$ with varying PFSS and SCS heights for 11-Jul-2010. The color bar indicates the SCS heights, crosses mark configurations that failed in the visual classification while dots mark configurations that passed it.}
    \label{Fig Topology 2010}
\end{figure*}

Using the full set of model parameters we apply, in the next step of our workflow, the topology classification for determining the match between open and closed magnetic field on the Sun, and calculate the three parameters $P_{\rm cov}$, $P_{\rm Jac}$ and $P_{\rm glob}$ (see Section \ref{sub:topology}). Figures~\ref{Fig Topology Maps 2008} and \ref{Fig Topology Maps} show the EUV Carrington maps for 01-Aug-2008 (Carrington rotation number 2072) and 11-Jul-2010 (Carrington rotation number 2098), respectively. The extracted coronal hole areas are overplotted together with the computed contours of EUHFORIA's open magnetic field. As can be seen for both dates, changing the boundary heights for $R_{\rm ss}$ and $R_{\rm scs}$, the computed open field area varies strongly and the lower these heights the more open field regions are generated. This is expected since lowering the $R_{\rm ss}$ height allows for more field lines to be considered by the model as open. The quantification of the overlap between modeled and observed open and closed field is given by the topology parameters described in the previous section.

Figure~\ref{Fig Topology 2008} and Figure~\ref{Fig Topology 2010} present the results from the different topology parameters together with the results from the visual classification. For both dates, we obtain that a lot of the highest scoring configurations from the topology analysis also passed the visual classification (indicated by large dots). For 01-Aug-2008 the three topology parameters behave differently, highlighting the different properties that the metrics measure. Results for $P_{\rm Jac}$ reveal an increase in match with increasing $R_{\rm ss}$, and consequently also with increasing $R_{\rm scs}$, up to a turning point at about 2.5~$R_\odot$. The parameter $P_{\rm cov}$ follows the expected trend of a continuous decrease with increasing configuration parameter values. This is due to the fact that with increasing $R_{\rm ss}$ and increasing $R_{\rm scs}$ less and less open field is generated by the model. Hence, the percentage of overlap of the model and observed open fields decreases. Inversely to $P_{\rm cov}$, the global parameter $P_{\rm glob}$ increases with increasing $R_{\rm ss}$ and $R_{\rm scs}$. 

For 11-Jul-2010, the results are slightly different as $P_{\rm glob}$ and $P_{\rm Jac}$ follow very similar trends, namely, increase if either $R_{\rm ss}$ or $R_{\rm scs}$ increases, with the latter one having the bigger impact. This means that there is a clear trend that modeling less open structures match better with the EUV observations for that date. $R_{\rm scs}$ is the dominating parameter here and could be used as limiting factor of open structures, because while the PFSS sets the magnetic topology, $R_{\rm scs}$ serves as the cut-off for the PFSS-domain and thus decides how much of the magnetic field is actually open. The model configuration that produces the lowest amount of open fields, having the highest $R_{\rm ss}$ and $R_{\rm scs}$ values, covers 3.88\% of the total area in comparison to the EUV observations giving about 3.00\%. The overestimation of open areas from the model for that date is also the reason for the inverse behavior of $P_{\rm cov}$ with respect to $P_{\rm glob}$ and $P_{\rm Jac}$, as $P_{\rm cov}$ is insensitive to overestimation and solely measures overlap regions between model and EUV. Hence, $P_{\rm Jac}$ should be used complementary to $P_{\rm cov}$ in order to derive the amount of overestimation of modeled open magnetic field areas.

The most general parameter we introduced here is $P_{\rm glob}$ which gives the fraction of matching pixels in the masks over the total number of pixels of the entire map. Therefore, $P_{\rm glob}$ reflects best the quality of the modelled output and is used as criterion for rejecting parameter sets of lower quality. We reduce the set of configurations by using the best 50\% from the distribution given by $P_{\rm glob}$, and subsequently, form set B (cf. Figure \ref{Fig: Workflow}). Set B consists, per definition of our criteria, of 33/33 parameter sets for 01-Aug-2008/11-Jul-2010. We note that $P_{\rm Jac}$ would give similar results, especially for the 11-Jul-2010 event. Interestingly, while there is a significant match between both the visual inspection step and the topology analysis, we can see in Figures \ref{Fig Topology 2008} and \ref{Fig Topology 2010} that the visual inspection would actually reject not only the worst configurations from the topology analysis, but also the best matching ones. Note that while for some configurations the general topology matches well, this does not necessarily mean that also the field line trajectories are matching when compared to white-light images. This implies that analyzing modelled open and closed fields yields additional information that cannot be derived through mere comparison of the field line configuration with white-light data and vice-versa.

\begin{figure*}
    \centering
    \includegraphics[scale = 0.4]{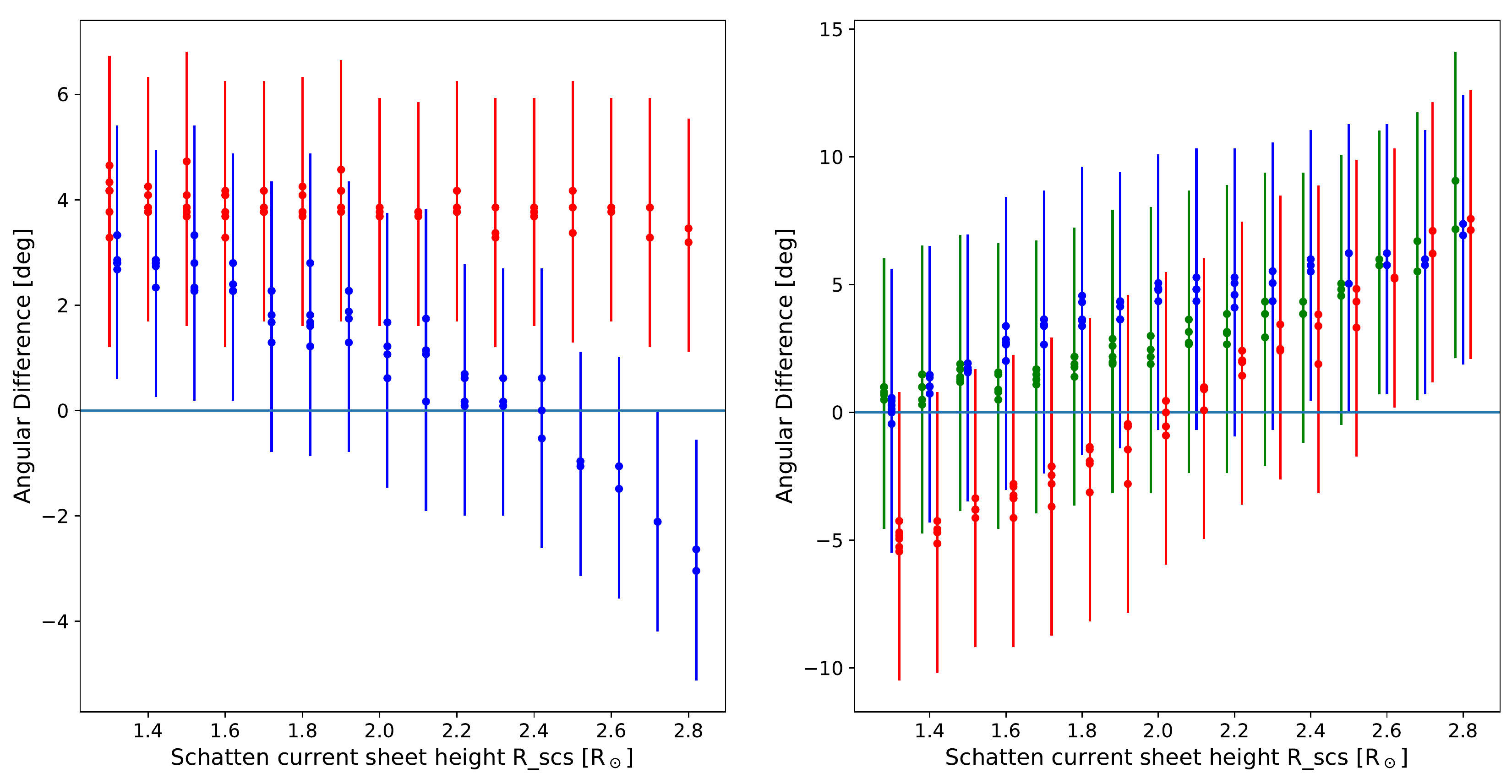}
    \caption{Difference of SCS field line angle and streamer angle from observations for 01-Aug-2008 (left) and 11-Jul-2010 (right). Red marks the results from SOHO/LASCO perspective, while blue and green are the results for STEREO/COR2. Multiple dots for a fixed SCS height indicate the different PFSS heights for the same $R_{\rm scs}$. }
    \label{FM1}
\end{figure*}

Starting again from the full set of model parameters we apply in the last step of the workflow the feature matching method (see Section \ref{sub:feature}). Using a semi-automatized algorithm we manually pick the brightest features, assuming that those lie closest to the plane-of-sky, in the white-light data and compare that with the modeled field lines (point-and-click method). Streamers can be characterized by their direction and width in the lower corona. Figure~\ref{FM1} shows for both dates in relation to the chosen SCS heights, and PFSS heights as given by multiple dots for the same $R_{\rm scs}$, the differences in the angles derived between modeled field line and observed streamer direction. Error bars reveal average uncertainties in the plane-of-sky selection for the SCS model results as these are 2D visualization slices. To define the error bars we simply varied the longitudinal direction by $+/-$10 degrees. We also investigated point-and-click inaccuracies but yielded a very minor effect compared to the errors as derived by the tilting procedure. For almost all chosen structures that we investigated, an approximately linear trend across the SCS height spectrum is obtained. The choice of the PFSS model boundary height seems to have a negligible influence on the resulting SCS field lines. One exception is the chosen feature from the LASCO perspective for 01-Aug-2008, where by using different model parameter sets no significant variation in the field line angle could be measured. As can be seen in Figure \ref{FM1}, for all the streamer directions there is no clear ideal SCS parameter derivable, most likely due to rather big errors originating from the uncertainty in the longitudinal streamer location itself.

\begin{figure*}
    \centering
    \includegraphics[scale = 0.4]{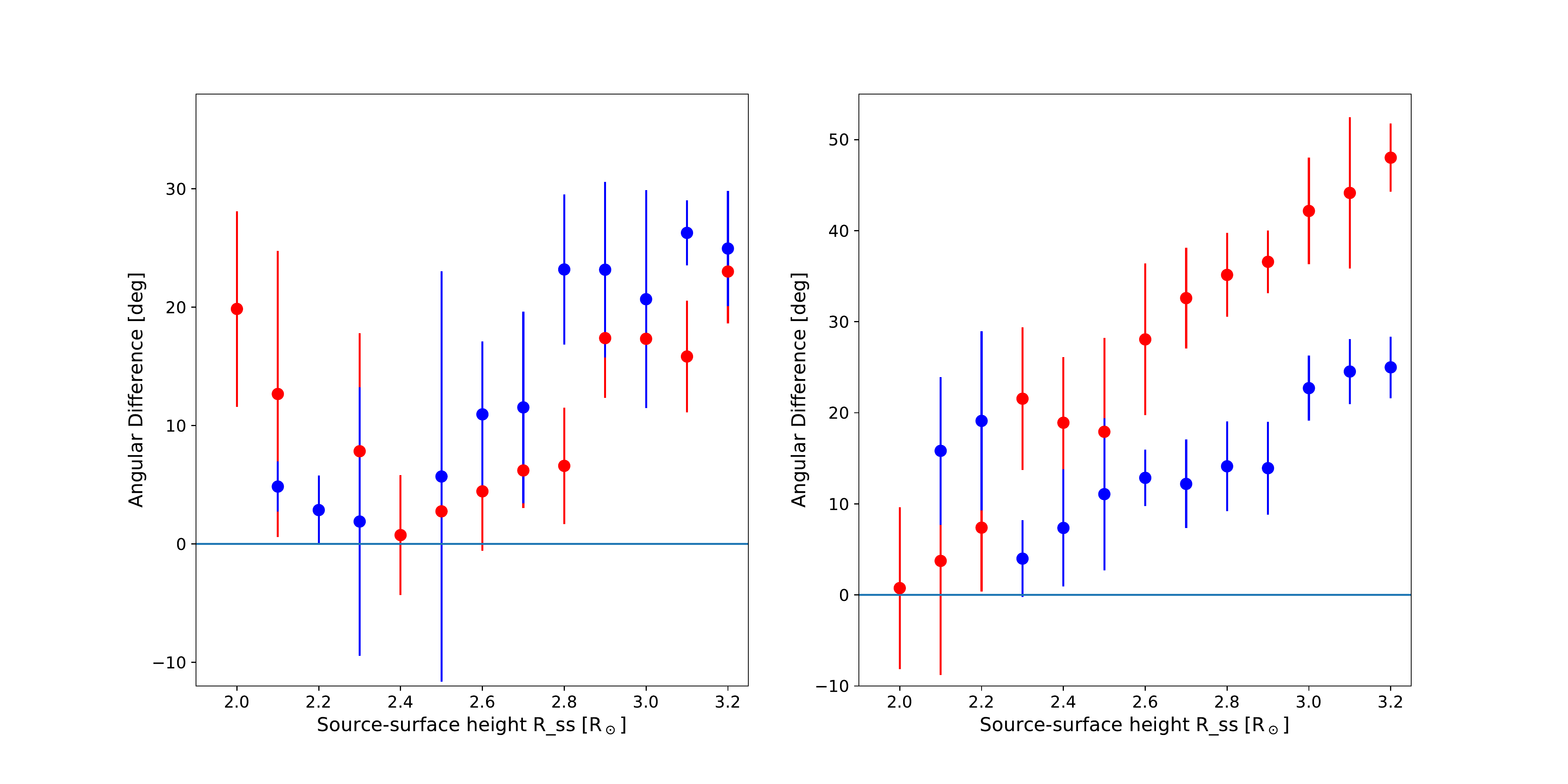}
    \caption{Difference of PFSS closed structure width and streamer width from observations for both 01-Aug-2008 (left) and 11-Jul-2010 (right). Red marks results originating from STEREO A, while shown in blue are the STEREO B results. }
    \label{FM2}
\end{figure*}

A more decisive picture is created by comparing the streamer width with the angular extension of underlying loop systems in the PFSS model. We derived the streamer widths from the white-light data as 34.8 and 18.6 degrees for the features selected for 01-Aug-2008 and 16.9 and 35.0 degrees for 11-Jul-2010 for Stereo A and B, respectively. The results are given in Figure~\ref{FM2} and we obtain that configurations with [$R_{\rm ss} \in {2.0 R_{\odot}; 2.5 R_{\odot}}$] seem to perform best for both dates. The errors originate from both the visualization bias, due to manually selecting which field lines are chosen to display, and the uncertainty due to the definition of the streamer width, either marking the closed loop system or the closest open field lines next to that closed system, as shown in Figure~\ref{Fig: FM}d). After extensive testing, we conclude that the point-and-click errors are comparatively small compared to these error sources.

\begin{figure}
    \centering
    \includegraphics[width=\linewidth]{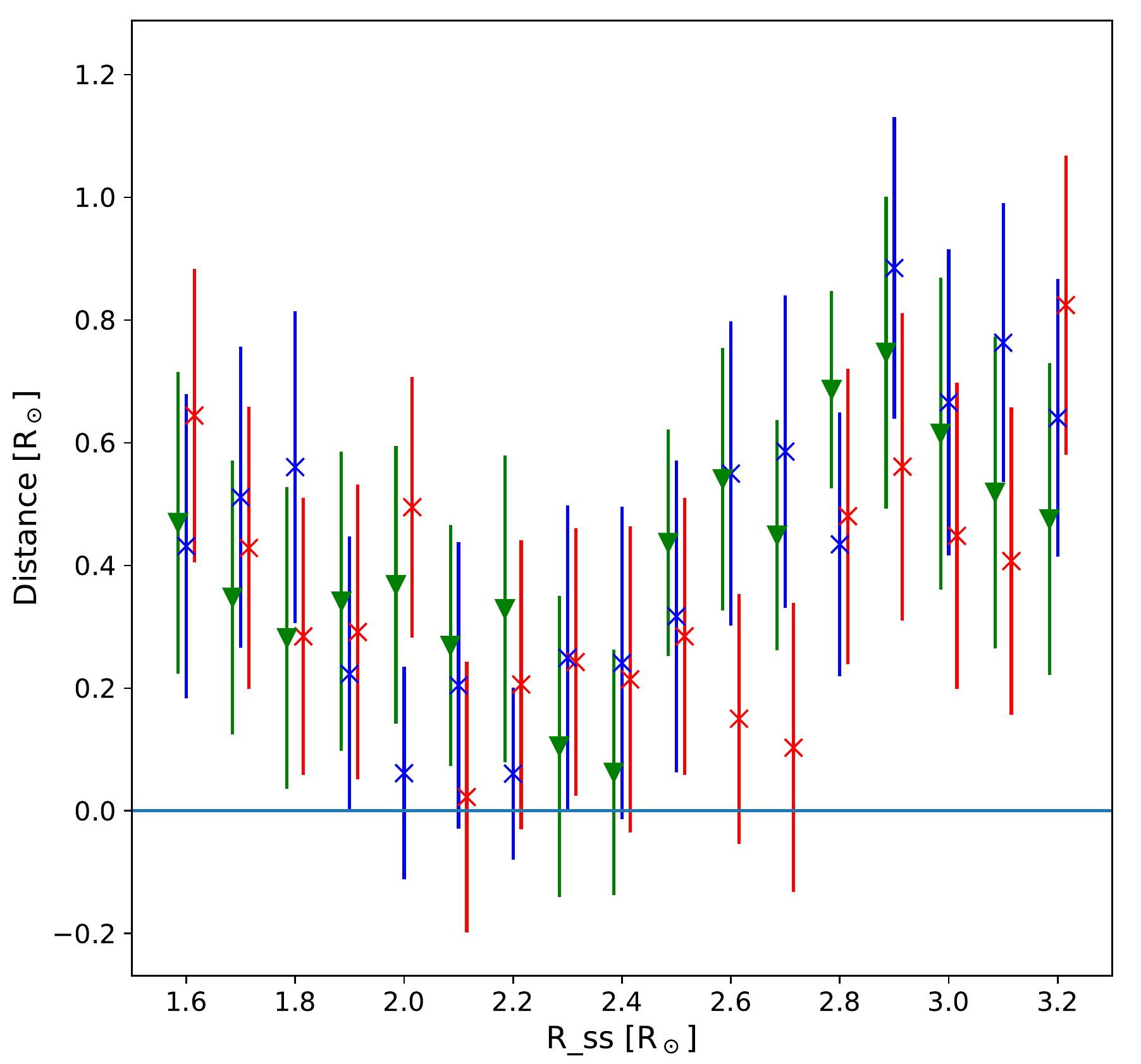}
    \caption{2D projected positional difference of PFSS loop structures and loop systems from eclipse observations by Druckmüller for both 08-01-2008 (crosses) and 07-11-2010 (triangles). Different colors indicate different features.}
    \label{FM3}
\end{figure}

The final routine we implemented involves the Brute Force feature matching to compare from PFSS results the location of prominent features such as the top of closed loop systems. As this requires a rather accurate identification of white-light structures for the comparison, high-resolution solar eclipse photographs are used for this analysis. Figure \ref{FM3} shows the results with errorbars coming from the visualization bias that the model results are imposed upon by the selection of a set of field lines. The lowest differences between model and observations are derived for configurations with $R_{ss} \in [2.0, 2.5] R_{\odot}$. Based on that we obtain a similar outcome as for the streamer width analysis (see Figure~\ref{FM2}).  

Results from the feature matching method form set C in our workflow. Set C covers configurations for which, for at least two out of three matched features, the error bars reach the line of zero difference to the observations, namely, the blue horizontal lines in Figures \ref{FM1}, \ref{FM2} and \ref{FM3}. It consists of 11/11 parameter sets for 01-Aug-2008/11-Jul-2010.

Inspecting the overlap in the reduced sets from A, B, and C (Table~\ref{conf_table}) we combine now the quality assessment of visual, topology and feature matching classifications. Interestingly, there exists one configuration that passed all of the three sub-steps for both dates. This configuration is $R_{\rm scs} = 2.0 R_{\odot}, R_{\rm ss} = 2.4 R_{\odot}$ and marks our derived ideal parameter set for this exemplary analysis.

\section{Discussion}
\label{sec5}

We developed in this study a validation scheme that acts as guideline for a standardized quality assessment of coronal model results. It presents a tool for modelers that can be easily applied in order to chose the most reliable option(s) of model input data and parameters among the many different possible ones. On the example of the EUHFORIA coronal model, we define classification steps based on comparing PFSS and SCS model results with observational data from different perspectives. The classification steps cover visual comparison of global open/closed magnetic field structures and isolated features as well as mathematical metrics, and can be used in an objective way to reduce the initial set of model configurations down to the most reliable ones.

For separating the initial set into good or bad matching model configurations we use a visual inspection focusing on the magnetic field line bending in the SCS model, close to the lower boundary, and compared that to coronagraph data from different instruments. We find that the visual comparison is subjective, but can be performed rather easily, and may quickly sort out a large parameter set. It is also a useful tool to investigate properties as e.g., the SCS bending, that cannot easily be determined by neither the topology parameters nor the feature matching classification.

An objective classification is given by the topology analysis, which uses the information of open magnetic field on the Sun as extracted from synoptic EUV image data. We perform a simple detection of dark areas in the EUV data, via an intensity threshold, that presumably represent coronal holes which are the dominant source of open magnetic field and fast solar wind from the Sun \citep[for a review see e.g.,][]{cranmer19}. We apply different parameters for comparing the model results with the observations such as $P_{\rm cov}$ \citep[to assess the performance of a model configuration with focus on open field computation, however, ignoring areas outside of EUV-open regions; see][]{Asvestari19}, $P_{\rm Jac}$ (to assess the amount of under- and overestimating the open field computation giving the percentage of similarity between the model- and the EUV-open regions), and $P_{\rm glob}$ (to assess the overall performance of specific model configurations, as it is the plain overlap percentage of both masks including closed regions). We find that $P_{\rm cov}$ and $P_{\rm Jac}$ values for 01-Aug-2008 are much lower compared to those from 11-Jul-2010 and the comparison is left to the $P_{\rm glob}$ parameter. We find that model configurations revealing a low percentage in the topology parameters were also rejected from the visual inspection. The boundary height of $R_{\rm scs}$ dominates the computed amount of open field and in general heights larger than $\sim$2~R$_{\odot}$ yield the better match to the observations. Uncertainties for that method definitely come from the assumption that open magnetic field predominantly originates from dark structures as observed in EUV and by using synoptic EUV data, representing the coronal structures over a full solar rotation (the same holds for the magnetic field input used for the model). The uncertainties in the dark area (i.e., coronal hole) extraction itself is found to lie in the range of $\pm$25\% \citep{linker21}. Taking that into account, still the coronal model is on the lower limit of matching with observations and generally overestimates the open magnetic field areas. In a recent paper by \cite{Asvestari19} it was shown that this has a negligible effect when using more coarse model resolutions of about 2 degrees per pixel. Moreover, EIT images (01-Aug-2008) seem to be more noisy compared to the AIA image data (11-Jul-2010) and the two dates under study cover different phases in the solar activity cycle with 2010 being a more active time compared to the minimum phase during 2008. High activity might cause large deviations from a steady-state condition and strong changes for the synoptic data which are used (as input and for the comparison). Nevertheless, the 2010 date produces numerically the better results in the topology analysis.

For the feature matching method we obtain for most of the results a strong dominance of $R_{\rm scs}$. Thus, the field line appearance of the SCS model only weakly depends on varying the source surface parameter $R_{\rm ss}$ of the underlying PFSS model. While the streamer angle comparison is found to be only a weak filter (a big portion of configurations pass with 33 of 67 for 01-Aug-2008 and 61 out of 67 for 11-Jul-2010), the situation is very different when comparing the width of the streamer and the Brute Force matching method. Applying those classifications we obtain an overall combined number of only 11 configurations left in set C for each date.

Our final conclusion from generating that benchmarking system on the example of EUHFORIA yields that the coronal model configuration with $R_{\rm scs} = 2.0 R_{\odot}, R_{\rm ss} = 2.4 R_{\odot}$ is the ideal parameter set for the analyzed dates. While a more in-depth analysis with a broader selection of dates would be necessary to draw a more comprehensive conclusion, this result is matching within the expected range in parameter space and conforms with current defaults/conventions \citep[see e.g.,][]{Mackay12,PomoellPoedts18}. 

The possible sequence of the classification steps as described are not depending on each other and can be applied in any desired order. For a large number of model configurations under investigation we may suggest combining methods A and B to reduce the parameter sets before further analysis. The strength of combining these two methods lies in the combination of empirical visual classification with a mathematical scheme.  

Input data and parameters for any coronal model underlie large variations and generate plentiful results that need to be assessed in quality and reliability with respect to observations. Standardized validation schemes, as presented here, are a necessity for model improvement leading to more reliable Space Weather forecasts \citep[see also][]{macneice18,Hinterreiter19,verbeke19}. Moreover, more reliable model results provide us a basis for better understanding the interplay between global open and closed magnetic field configuration resulting in the different solar wind structures, which in turn leads to a better understanding of the propagation characteristics of coronal mass ejections in interplanetary space.

\begin{acknowledgements} 
A.W. acknowledges financial support by the CCSOM project. E.A. acknowledges the support of the Academy of Finland (Project TRAMSEP, Academy of Finland Grant 322455).
\end{acknowledgements}

\bibliographystyle{aa}

\begin{thebibliography}{48}
\expandafter\ifx\csname natexlab\endcsname\relax\def\natexlab#1{#1}\fi

\bibitem[{{Arge} \& {Pizzo}(2000)}]{Arge2000}
{Arge}, C.~N. \& {Pizzo}, V.~J. 2000, \jgr, 105, 10465

\bibitem[{{Asvestari} {et~al.}(2019){Asvestari}, {Heinemann}, {Temmer},
  {Pomoell}, {Kilpua}, {Magdalenic}, \& {Poedts}}]{Asvestari19}
{Asvestari}, E., {Heinemann}, S.~G., {Temmer}, M., {et~al.} 2019, Journal of
  Geophysical Research (Space Physics), 124, 8280

\bibitem[{{Brueckner} {et~al.}(1995){Brueckner}, {Howard}, {Koomen},
  {Korendyke}, {Michels}, {Moses}, {Socker}, {Dere}, {Lamy}, {Llebaria},
  {Bout}, {Schwenn}, {Simnett}, {Bedford}, \& {Eyles}}]{Brueckner1995}
{Brueckner}, G.~E., {Howard}, R.~A., {Koomen}, M.~J., {et~al.} 1995, \solphys,
  162, 357

\bibitem[{{Case} {et~al.}(2008){Case}, {Spence}, {Owens}, {Riley}, \&
  {Odstrcil}}]{case08}
{Case}, A.~W., {Spence}, H.~E., {Owens}, M.~J., {Riley}, P., \& {Odstrcil}, D.
  2008, \grl, 35, L15105

\bibitem[{Childs {et~al.}(2012)Childs, Brugger, Whitlock, Meredith, Ahern,
  Pugmire, Biagas, Miller, Harrison, Weber, Krishnan, Fogal, Sanderson, Garth,
  Bethel, Camp, R\"{u}bel, Durant, Favre, \& Navr\'{a}til}]{VisIt}
Childs, H., Brugger, E., Whitlock, B., {et~al.} 2012, in {High Performance
  Visualization--Enabling Extreme-Scale Scientific Insight}, 357--372

\bibitem[{{Cohen} {et~al.}(2007){Cohen}, {Sokolov}, {Roussev}, \&
  {Gombosi}}]{Cohen07}
{Cohen}, O., {Sokolov}, I.~V., {Roussev}, I.~I., \& {Gombosi}, T.~I. 2007, in
  AGU Fall Meeting Abstracts, Vol. 2007, SH51B--06

\bibitem[{{Couvidat} {et~al.}(2016){Couvidat}, {Schou}, {Hoeksema}, {Bogart},
  {Bush}, {Duvall}, {Liu}, {Norton}, \& {Scherrer}}]{Couvidat2016}
{Couvidat}, S., {Schou}, J., {Hoeksema}, J.~T., {et~al.} 2016, \solphys, 291,
  1887

\bibitem[{{Cranmer}(2009)}]{Cranmer2009}
{Cranmer}, S.~R. 2009, Living Reviews in Solar Physics, 6, 3

\bibitem[{{Cranmer} {et~al.}(2017){Cranmer}, {Gibson}, \& {Riley}}]{cranmer17}
{Cranmer}, S.~R., {Gibson}, S.~E., \& {Riley}, P. 2017, Space Science Reviews,
  212, 1345

\bibitem[{{Cranmer} \& {Winebarger}(2019)}]{cranmer19}
{Cranmer}, S.~R. \& {Winebarger}, A.~R. 2019, \araa, 57, 157

\bibitem[{{Domingo} {et~al.}(1995){Domingo}, {Fleck}, \&
  {Poland}}]{Domingo1995}
{Domingo}, V., {Fleck}, B., \& {Poland}, A.~I. 1995, \solphys, 162, 1

\bibitem[{{Druckm{\"u}ller}(2009)}]{druckmueller09}
{Druckm{\"u}ller}, M. 2009, \apj, 706, 1605

\bibitem[{{Druckm{\"u}ller} {et~al.}(2006){Druckm{\"u}ller}, {Ru{\v{s}}in}, \&
  {Minarovjech}}]{Druckmueller06}
{Druckm{\"u}ller}, M., {Ru{\v{s}}in}, V., \& {Minarovjech}, M. 2006,
  Contributions of the Astronomical Observatory Skalnate Pleso, 36, 131

\bibitem[{{Harvey} {et~al.}(1996){Harvey}, {Hill}, {Hubbard}, {Kennedy},
  {Leibacher}, {Pintar}, {Gilman}, {Noyes}, {Title}, {Toomre}, {Ulrich},
  {Bhatnagar}, {Kennewell}, {Marquette}, {Patron}, {Saa}, \&
  {Yasukawa}}]{Harvey1996}
{Harvey}, J.~W., {Hill}, F., {Hubbard}, R.~P., {et~al.} 1996, Science, 272,
  1284

\bibitem[{{Heinemann} {et~al.}(2021){Heinemann}, {Saqri}, {Veronig},
  {Hofmeister}, \& {Temmer}}]{Heinemann2021}
{Heinemann}, S.~G., {Saqri}, J., {Veronig}, A.~M., {Hofmeister}, S.~J., \&
  {Temmer}, M. 2021, \solphys, 296, 18

\bibitem[{{Hess Webber} {et~al.}(2014){Hess Webber}, {Karna}, {Pesnell}, \&
  {Kirk}}]{HessWebber14}
{Hess Webber}, S.~A., {Karna}, N., {Pesnell}, W.~D., \& {Kirk}, M.~S. 2014,
  \solphys, 289, 4047

\bibitem[{{Hinterreiter} {et~al.}(2019){Hinterreiter}, {Magdalenic}, {Temmer},
  {Verbeke}, {Jebaraj}, {Samara}, {Asvestari}, {Poedts}, {Pomoell}, {Kilpua},
  {Rodriguez}, {Scolini}, \& {Isavnin}}]{Hinterreiter19}
{Hinterreiter}, J., {Magdalenic}, J., {Temmer}, M., {et~al.} 2019, \solphys,
  294, 170

\bibitem[{{Howard} {et~al.}(2002){Howard}, {Moses}, {Socker}, {Dere}, {Cook},
  \& {Secchi Consortium}}]{Howard2002}
{Howard}, R.~A., {Moses}, J.~D., {Socker}, D.~G., {et~al.} 2002, Advances in
  Space Research, 29, 2017

\bibitem[{{Howard} {et~al.}(2008){Howard}, {Moses}, {Vourlidas}, {Newmark},
  {Socker}, {Plunkett}, {Korendyke}, {Cook}, {Hurley}, {Davila}, {Thompson},
  {St Cyr}, {Mentzell}, {Mehalick}, {Lemen}, {Wuelser}, {Duncan}, {Tarbell},
  {Wolfson}, {Moore}, {Harrison}, {Waltham}, {Lang}, {Davis}, {Eyles},
  {Mapson-Menard}, {Simnett}, {Halain}, {Defise}, {Mazy}, {Rochus}, {Mercier},
  {Ravet}, {Delmotte}, {Auchere}, {Delaboudiniere}, {Bothmer}, {Deutsch},
  {Wang}, {Rich}, {Cooper}, {Stephens}, {Maahs}, {Baugh}, {McMullin}, \&
  {Carter}}]{Howard08}
{Howard}, R.~A., {Moses}, J.~D., {Vourlidas}, A., {et~al.} 2008, \ssr, 136, 67

\bibitem[{{Jian} {et~al.}(2016){Jian}, {MacNeice}, {Mays}, {Taktakishvili},
  {Odstrcil}, {Jackson}, {Yu}, {Riley}, \& {Sokolov}}]{Jian16}
{Jian}, L.~K., {MacNeice}, P.~J., {Mays}, M.~L., {et~al.} 2016, Space Weather,
  14, 592

\bibitem[{{Kaiser}(2005)}]{Kaiser2005}
{Kaiser}, M.~L. 2005, Advances in Space Research, 36, 1483

\bibitem[{{Kaiser} {et~al.}(2008){Kaiser}, {Kucera}, {Davila}, {St. Cyr},
  {Guhathakurta}, \& {Christian}}]{Kaiser08}
{Kaiser}, M.~L., {Kucera}, T.~A., {Davila}, J.~M., {et~al.} 2008, \ssr, 136, 5

\bibitem[{{Karna} {et~al.}(2014){Karna}, {Hess Webber}, \& {Pesnell}}]{Karna14}
{Karna}, N., {Hess Webber}, S.~A., \& {Pesnell}, W.~D. 2014, \solphys, 289,
  3381

\bibitem[{{Koutchmy} \& {Livshits}(1992)}]{koutchmy92}
{Koutchmy}, S. \& {Livshits}, M. 1992, \ssr, 61, 393

\bibitem[{{Krista} \& {Gallagher}(2009)}]{krista09}
{Krista}, L.~D. \& {Gallagher}, P.~T. 2009, \solphys, 256, 87

\bibitem[{{Linker} {et~al.}(2021){Linker}, {Heinemann}, {Temmer}, {Owens},
  {Caplan}, {Arge}, {Asvestari}, {Delouille}, {Downs}, {Hofmeister}, {Jebaraj},
  {Madjarska}, {Pinto}, {Pomoell}, {Samara}, {Scolini}, \&
  {Vr{\v{s}}nak}}]{linker21}
{Linker}, J.~A., {Heinemann}, S.~G., {Temmer}, M., {et~al.} 2021, \apj, 918, 21

\bibitem[{{Mackay} \& {Yeates}(2012)}]{Mackay12}
{Mackay}, D.~H. \& {Yeates}, A.~R. 2012, Living Reviews in Solar Physics, 9, 6

\bibitem[{{MacNeice} {et~al.}(2018){MacNeice}, {Jian}, {Antiochos}, {Arge},
  {Bussy-Virat}, {DeRosa}, {Jackson}, {Linker}, {Mikic}, {Owens}, {Ridley},
  {Riley}, {Savani}, \& {Sokolov}}]{macneice18}
{MacNeice}, P., {Jian}, L.~K., {Antiochos}, S.~K., {et~al.} 2018, Space
  Weather, 16, 1644

\bibitem[{{McGregor} {et~al.}(2008){McGregor}, {Hughes}, {Arge}, \&
  {Owens}}]{mcgregor08}
{McGregor}, S.~L., {Hughes}, W.~J., {Arge}, C.~N., \& {Owens}, M.~J. 2008,
  Journal of Geophysical Research (Space Physics), 113, A08112

\bibitem[{{Meyer} {et~al.}(2020){Meyer}, {Mackay}, {Talpeanu}, {Upton}, \&
  {West}}]{Meyer20}
{Meyer}, K.~A., {Mackay}, D.~H., {Talpeanu}, D.-C., {Upton}, L.~A., \& {West},
  M.~J. 2020, \solphys, 295, 101

\bibitem[{{Morgan} {et~al.}(2006){Morgan}, {Habbal}, \& {Woo}}]{Morgan06}
{Morgan}, H., {Habbal}, S.~R., \& {Woo}, R. 2006, Solar Physics, 236, 263

\bibitem[{{Pesnell} {et~al.}(2012){Pesnell}, {Thompson}, \&
  {Chamberlin}}]{Pesnell12}
{Pesnell}, W.~D., {Thompson}, B.~J., \& {Chamberlin}, P.~C. 2012, \solphys,
  275, 3

\bibitem[{{Pomoell} \& {Poedts}(2018)}]{PomoellPoedts18}
{Pomoell}, J. \& {Poedts}, S. 2018, Journal of Space Weather and Space Climate,
  8, A35

\bibitem[{{Riley} {et~al.}(2014){Riley}, {Ben-Nun}, {Linker}, {Mikic},
  {Svalgaard}, {Harvey}, {Bertello}, {Hoeksema}, {Liu}, \&
  {Ulrich}}]{Riley2014}
{Riley}, P., {Ben-Nun}, M., {Linker}, J.~A., {et~al.} 2014, \solphys, 289, 769

\bibitem[{{Riley} {et~al.}(2018){Riley}, {Mays}, {Andries}, {Amerstorfer},
  {Biesecker}, {Delouille}, {Dumbovi{\'c}}, {Feng}, {Henley}, {Linker},
  {M{\"o}stl}, {Nu{\~n}ez}, {Pizzo}, {Temmer}, {Tobiska}, {Verbeke}, {West}, \&
  {Zhao}}]{riley18}
{Riley}, P., {Mays}, M.~L., {Andries}, J., {et~al.} 2018, Space Weather, 16,
  1245

\bibitem[{{Rotter} {et~al.}(2012){Rotter}, {Veronig}, {Temmer}, \&
  {Vr{\v{s}}nak}}]{rotter12}
{Rotter}, T., {Veronig}, A.~M., {Temmer}, M., \& {Vr{\v{s}}nak}, B. 2012,
  \solphys, 281, 793

\bibitem[{{Sachdeva} {et~al.}(2015){Sachdeva}, {Subramanian}, {Colaninno}, \&
  {Vourlidas}}]{sachdeva15}
{Sachdeva}, N., {Subramanian}, P., {Colaninno}, R., \& {Vourlidas}, A. 2015,
  \apj, 809, 158

\bibitem[{{Sasso} {et~al.}(2019){Sasso}, {Pinto}, {Andretta}, {Howard},
  {Vourlidas}, {Bemporad}, {Dolei}, {Spadaro}, {Susino}, {Antonucci}, {Abbo},
  {Da Deppo}, {Fineschi}, {Frassetto}, {Landini}, {Naletto}, {Nicolini},
  {Nicolosi}, {Pancrazzi}, {Romoli}, {Telloni}, \& {Ventura}}]{Sasso19}
{Sasso}, C., {Pinto}, R.~F., {Andretta}, V., {et~al.} 2019, \aap, 627, A9

\bibitem[{{Schatten}(1971)}]{Schatten1971}
{Schatten}, K.~H. 1971, Cosmic Electrodynamics, 2, 232

\bibitem[{{Schmidt} \& {Cargill}(2001)}]{schmidt01}
{Schmidt}, J.~M. \& {Cargill}, P.~J. 2001, \jgr, 106, 8283

\bibitem[{{Schou} {et~al.}(2012){Schou}, {Scherrer}, {Bush}, {Wachter},
  {Couvidat}, {Rabello-Soares}, {Bogart}, {Hoeksema}, {Liu}, {Duvall}, {Akin},
  {Allard}, {Miles}, {Rairden}, {Shine}, {Tarbell}, {Title}, {Wolfson},
  {Elmore}, {Norton}, \& {Tomczyk}}]{Schou2012}
{Schou}, J., {Scherrer}, P.~H., {Bush}, R.~I., {et~al.} 2012, \solphys, 275,
  229

\bibitem[{{Sheeley} {et~al.}(1997){Sheeley}, {Wang}, {Hawley}, {Brueckner},
  {Dere}, {Howard}, {Koomen}, {Korendyke}, {Michels}, {Paswaters}, {Socker},
  {St. Cyr}, {Wang}, {Lamy}, {Llebaria}, {Schwenn}, {Simnett}, {Plunkett}, \&
  {Biesecker}}]{sheeley97}
{Sheeley}, N.~R., {Wang}, Y.~M., {Hawley}, S.~H., {et~al.} 1997, \apj, 484, 472

\bibitem[{{Temmer}(2021)}]{temmer21}
{Temmer}, M. 2021, Living Reviews in Solar Physics, 18, 4

\bibitem[{{Temmer} {et~al.}(2011){Temmer}, {Rollett}, {M{\"o}stl}, {Veronig},
  {Vr{\v{s}}nak}, \& {Odstr{\v{c}}il}}]{temmer11}
{Temmer}, M., {Rollett}, T., {M{\"o}stl}, C., {et~al.} 2011, \apj, 743, 101

\bibitem[{{Verbeke} {et~al.}(2019){Verbeke}, {Mays}, {Temmer}, {Bingham},
  {Steenburgh}, {Dumbovi{\'c}}, {N{\'u}{\~n}ez}, {Jian}, {Hess}, {Wiegand},
  {Taktakishvili}, \& {Andries}}]{verbeke19}
{Verbeke}, C., {Mays}, M.~L., {Temmer}, M., {et~al.} 2019, Space Weather, 17, 6

\bibitem[{{Vr{\v{s}}nak}(2021)}]{vrsnak21}
{Vr{\v{s}}nak}, B. 2021, Journal of Space Weather and Space Climate, 11, 34

\bibitem[{{Wang} {et~al.}(2007){Wang}, {Sheeley}, \& {Rich}}]{wang07}
{Wang}, Y.~M., {Sheeley}, N.~R., J., \& {Rich}, N.~B. 2007, \apj, 658, 1340

\bibitem[{{Yeates} {et~al.}(2018){Yeates}, {Amari}, {Contopoulos}, {Feng},
  {Mackay}, {Miki{\'c}}, {Wiegelmann}, {Hutton}, {Lowder}, {Morgan}, {Petrie},
  {Rachmeler}, {Upton}, {Canou}, {Chopin}, {Downs}, {Druckm{\"u}ller},
  {Linker}, {Seaton}, \& {T{\"o}r{\"o}k}}]{Yeates18}
{Yeates}, A.~R., {Amari}, T., {Contopoulos}, I., {et~al.} 2018, \ssr, 214, 99

\end{thebibliography}

\begin{appendix}
\section{Full list of model parameter sets}
Table\ref{conf_table} lists the full set of model parameters that were used for producing 67 different configurations in the EUHFORIA coronal model results as used for the analysis given in Section~\ref{sec4}. In addition, for each date (2008 and 2010) we give the parameter sets that passed (x) the criteria for visual (A), topology (B), or feature matching (C) classification. 

\begin{center}
\begin{longtable}{||c | c | c | c | c | c | c | c | c ||} 
 \hline
 Configuration Number & $R_{\mathrm{scs}} [\mathrm{R_\odot}]$ & $R_{\mathrm{ss}} [\mathrm{R_\odot}]$ & 2008 A & 2008 B & 2008 C & 2010 A & 2010 B & 2010 C \\ 
 \hline
 1 &  1.3 & 1.4 & -- & -- & -- & -- & -- & -- \\ 
 \hline
 2 & 1.3 & 1.7 & -- & -- & -- & -- & -- & -- \\
 \hline
 3 & 1.3 & 2.0 & -- & -- & -- & -- & -- & x \\
 \hline
 4 & 1.3 & 2.3 & -- & -- & -- & -- & -- & x \\
 \hline
 5 & 1.3 & 2.6 & -- & -- & -- & -- & -- & -- \\ 
 \hline
 6 &  1.3 & 2.9 & -- & -- & -- & -- & -- & -- \\ 
 \hline
 7 & 1.3 & 3.2 & -- & -- & -- & -- & -- & -- \\
 \hline
 8 & 1.4 & 1.5 & -- & -- & -- & -- & -- & -- \\
 \hline
 9 & 1.4 & 1.8 & -- & -- & -- & -- & -- & -- \\
 \hline
 10 & 1.4 & 2.1 & -- & -- & -- & -- & -- & -- \\
  \hline
 11 & 1.4 & 2.4 & -- & -- & -- & -- & -- & x \\
  \hline
 12 & 1.4 & 2.7 & -- & -- & -- & -- & -- & -- \\
  \hline
 13 & 1.4 & 3.0 & -- & -- & -- & -- & -- & -- \\
  \hline
 14 & 1.5 & 1.6 & -- & -- & -- & -- & -- & -- \\
  \hline
 15 & 1.5 & 1.9 & -- & -- & -- & x & -- & -- \\
   \hline
 16 & 1.5 & 2.2 & x & -- & -- & x & -- & -- \\
  \hline
 17 & 1.5 & 2.5 & x & -- & -- & x & -- & -- \\
   \hline
 18 & 1.5 & 2.8 & x & -- & -- & x & -- & -- \\
  \hline
 19 & 1.5 & 3.1 & -- & -- & -- & x & -- & -- \\
   \hline
 20 & 1.6 & 1.7 & -- & -- & -- & -- & -- & -- \\
  \hline
 21 & 1.6 & 2.0 & x & -- & -- & x & -- & x \\
    \hline
 22 & 1.6 & 2.3 & x & -- & -- & x & -- & x \\
  \hline
 23 & 1.6 & 2.6 & x & -- & -- & x & -- & -- \\
    \hline
 24 & 1.6 & 2.9 & x & -- & -- & x & -- & -- \\
  \hline
 25 & 1.6 & 3.2 & x & -- & -- & x & -- & -- \\
    \hline
 26 & 1.7 & 1.8 & x & -- & -- & x & -- & -- \\
  \hline
 27 & 1.7 & 2.1 & x & -- & x & x & -- & --\\
   \hline
 28 & 1.7 & 2.4 & x & -- & -- & x & -- & x \\
  \hline
 29 & 1.7 & 2.7 & x & x & -- & x & -- & -- \\
     \hline
 30 & 1.7 & 3.0 & x & x & -- & x & x & -- \\
  \hline
 31 & 1.8 & 1.9 & x & -- & -- & x & -- & -- \\
   \hline
 32 & 1.8 & 2.2 & x & -- & x & x & -- & --\\
   \hline
 33 & 1.8 & 2.5 & x & x & x & x & x & -- \\
   \hline
 34 & 1.8 & 2.8 & x & x & -- & x & x & -- \\
   \hline
 35 & 1.8 & 3.1 & x & x & -- & x & x & -- \\
   \hline
 36 & 1.9 & 2.0 & x & -- & -- & x & -- & x \\
    \hline
 37 & 1.9 & 2.3 & x & -- & x & x & x & x \\
    \hline
 38 & 1.9 & 2.6 & x & x & -- & x & x & -- \\
    \hline
 39 & 1.9 & 2.9 & x & x & -- & x & x & -- \\
    \hline
 40 & 1.9 & 3.2 & x & x & -- & x & x & -- \\
     \hline
 41 & 2.0 & 2.1 & x & -- & x & x & -- & -- \\
     \hline
 42 & 2.0 & 2.4 & x & x & x & x & x & x \\
     \hline
 43 & 2.0 & 2.7 & x & x & -- & x & x & -- \\
     \hline
 44 & 2.0 & 3.0 & -- & x & -- & x & x & -- \\
      \hline
 45 & 2.1 & 2.2 & x & -- & x & x & -- & -- \\
      \hline
 46 & 2.1 & 2.5 & x & x & x & x & x & -- \\
       \hline
 47 & 2.1 & 2.8 & x & x & -- & x & x & -- \\
       \hline
 48 & 2.1 & 3.1 & -- & x & -- & -- & x & -- \\
       \hline
 49 & 2.2 & 2.3 & -- & x & x & -- & x & x \\
        \hline
 50 & 2.2 & 2.6 & x & x & -- & -- & x & -- \\
        \hline
 51 & 2.2 & 2.9 & -- & x & -- & -- & x & -- \\
        \hline
 52 & 2.2 & 3.2 & -- & x & -- & -- & x & -- \\
        \hline
 53 & 2.3 & 2.4 & -- & x & x & -- & x & x \\
         \hline
 54 & 2.3 & 2.7 & -- & x & -- & -- & x & -- \\
         \hline
 55 & 2.3 & 3.0 & -- & x & -- & -- & x & -- \\
         \hline
 56 & 2.4 & 2.5 & -- & x & x & -- & x & -- \\
         \hline
 57 & 2.4 & 2.8 & -- & x & -- & -- & x & -- \\
         \hline
 58 & 2.4 & 3.1 & -- & x & -- & -- & x & -- \\
         \hline
 59 & 2.5 & 2.6 & -- & x & -- & -- & x & -- \\
         \hline
 60 & 2.5 & 2.9 & -- & x & -- & -- & x & -- \\
         \hline
 61 & 2.5 & 3.2 & -- & x & -- & -- & x & -- \\
          \hline
 62 & 2.6 & 2.7 & -- & x & -- & -- & x & -- \\
          \hline
 63 & 2.6 & 3.0 & -- & x & -- & -- & x & -- \\
          \hline
 64 & 2.7 & 2.8 & -- & x & -- & -- & x & -- \\
          \hline
 65 & 2.7 & 3.1 & -- & x & -- & -- & x & -- \\
          \hline
 66 & 2.8 & 2.9 & -- & x & -- & -- & x & -- \\
          \hline
 67 & 2.8 & 3.2 & -- & x & -- & -- & x & -- \\
  \hline
   \caption{Used parameter sets with the PFSS and SCS model heights for the analysis as described in Section~\ref{sec4}. Those parameter sets that passed the criteria for high quality in the visual (A), topology (B), or feature matching (C) classification scheme are marked by x. }
   \label{conf_table}
\end{longtable}
\end{center}

\end{appendix}

\end{document}